# Codon Capture and Ambiguous Intermediate Scenarios of Genetic Code Evolution

Tatsuro YAMASHITA and Osamu NARIKIYO

*Department of Physics, Kyushu University,*
*Hakozaki, Higashi-ku, Fukuoka 812-8581*

Using the shape space of codons and tRNAs we give a physical description of the genetic code evolution on the basis of the codon capture and ambiguous intermediate scenarios in a consistent manner. In the lowest dimensional version of our description, a physical quantity, codon level is introduced. In terms of the codon levels two scenarios are typically classified into two different routes of the evolutional process. In the case of the ambiguous intermediate scenario we perform an evolutional simulation implemented cost selection of amino acids and confirm a rapid transition of the code change. Such rapidness reduces uncomfortableness of the non-unique translation of the code at intermediate state that is the weakness of the scenario. In the case of the codon capture scenario the survival against mutations under the mutational pressure minimizing GC content in genomes is simulated and it is demonstrated that cells which experience only neutral mutations survive.

## 1. Introduction

The genetic code is the central element of every biological phenomenon. It was supposed to be frozen at first, but it turned out to be evolvable after the discovery of nonstandard code in mitochondria.[1)] After that several scenarios for the evolution of the genetic code,[1–3)] namely the reassignment of codons to amino acids, have been proposed. Among them we focus on the codon capture[1,4)] and ambiguous intermediate[5,6)] scenarios along preceding studies.[3,7,8)]

In the text-book argument[1)] the possibility of the latter scenario was questioned. However, recent analysis[7)] for observed codon reassignments supports both scenarios according to the situation. The codon capture scenario is supported by observed codon reassignments from stop to an amino acid. On the other hand, the ambiguous intermediate scenario is supported by observed codon reassignments from one amino acid to another.

In order to give a unified description of these two scenarios we try to construct a physical model by making use of the shape space of codons and tRNAs where the molecular recognition between them plays a central role. Most reported studies[9–12)] of the code evolution concern the sequences of codons and anticodons. A unique study[13)] on the basis of a chemical reaction network has also been reported. In contrast our study is based on the shape space after the study of immunity[14)] where a shape space is employed to describe the molecular recognition between antigens and antibodies. On the basis of our physical model we can simulate the ambiguous intermediate scenario in the case where the cost selection[7)] mechanism works and the codon capture scenario in the case only neutral mutations are accepted[1)] by the natural selection. Our simulation in the shape space illustrates the evolutional process in a chronological manner. Theoretical controversy between the advocates of two scenarios is naturally resolved from the view point of shape space.

This paper is organized as follows. In § 2 we summarize two scenarios and introduce our physical model. Some simulation results on the basis of our model are discussed in § 3 focusing on the time dependence of the genetic code evolution. Our conclusion is given in § 4.

## 2. Model

2.1. Codon capture and ambiguous intermediate scenarios

The abstract of the codon capture[1,4] and ambiguous intermediate[5,6] scenarios are shown in Fig. 1. The codon reassignments arise from alterations in tRNA, aaRS, ribosome, release factor and so on. Here in this paper we focus on the change in the function of tRNA as the trigger of the code evolution. Such a simplification might lack a reality but we do not quantitatively evaluate the relative importance among these alterations so that we choose tRNA as a representative factor to implement the variation in the correspondence between codons and amino-acids in the most direct way. The alterations in the other factors, which have been neglected for simplicity, should be included to obtain a realistic model. According to the position of the tRNA alteration, favored scenario changes as shown in Fig. 2. The typical situation is summarized[2,3] as follows and we adopt this summary as a basis of our model. If the alteration occurs at the anticodon in tRNA, the codon reassignment proceeds according as the codon capture scenario. On the other hand, if the alteration occurs at locations other than the anticodon in tRNA, the codon reassignment proceeds according to the ambiguous intermediate scenario. The alteration is caused by mutation of bases, base modification, RNA editing and so on. In the following we do not specify the cause of the alteration and only pay attention to the resulting change in the function of tRNA.

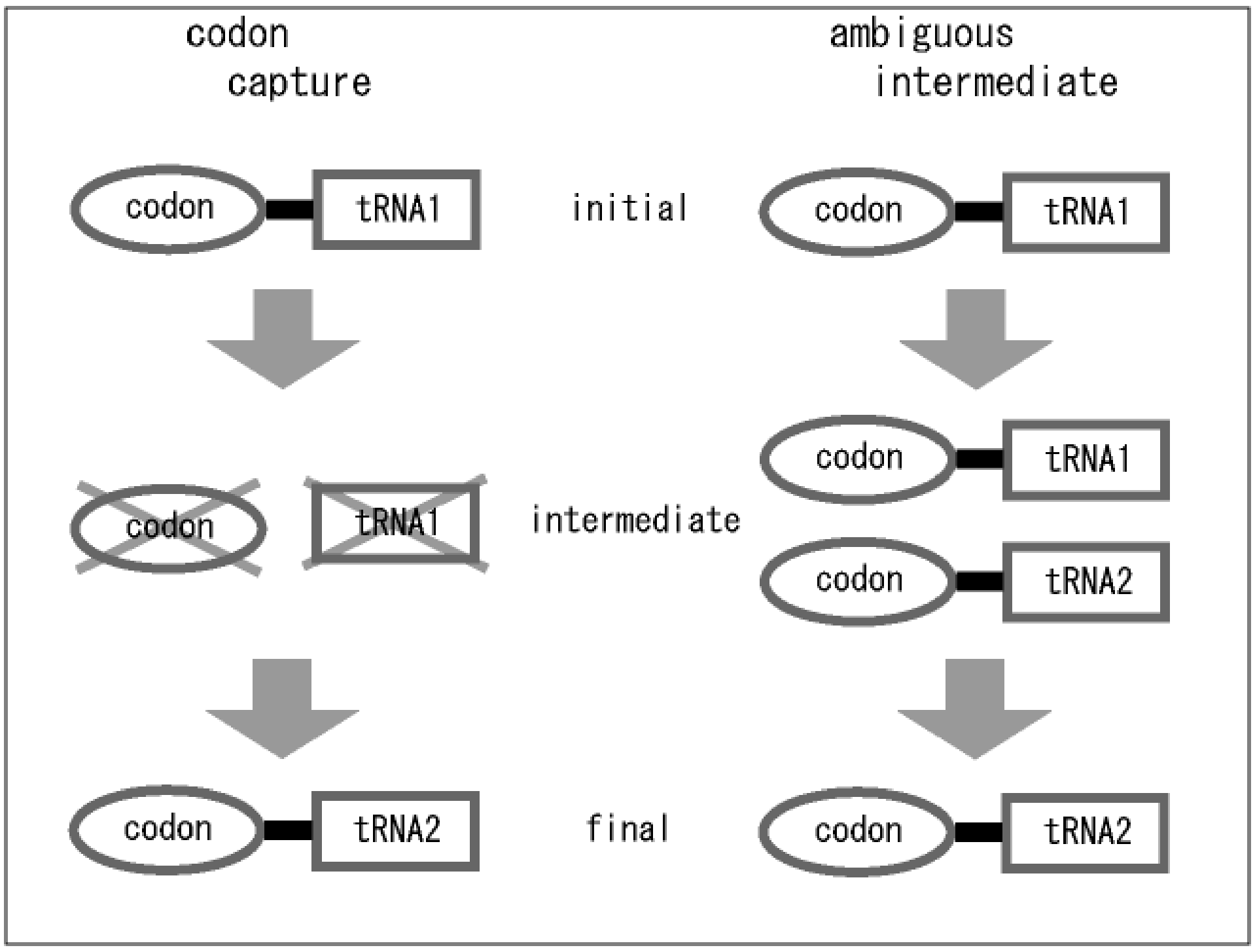

**Fig. 1.** Codon capture and ambiguous intermediate scenarios for the genetic code evolution. In the intermediate state both codon and tRNA disappear in codon capture scenario but a codon is recognized by two different tRNAs in ambiguous intermediate scenario.

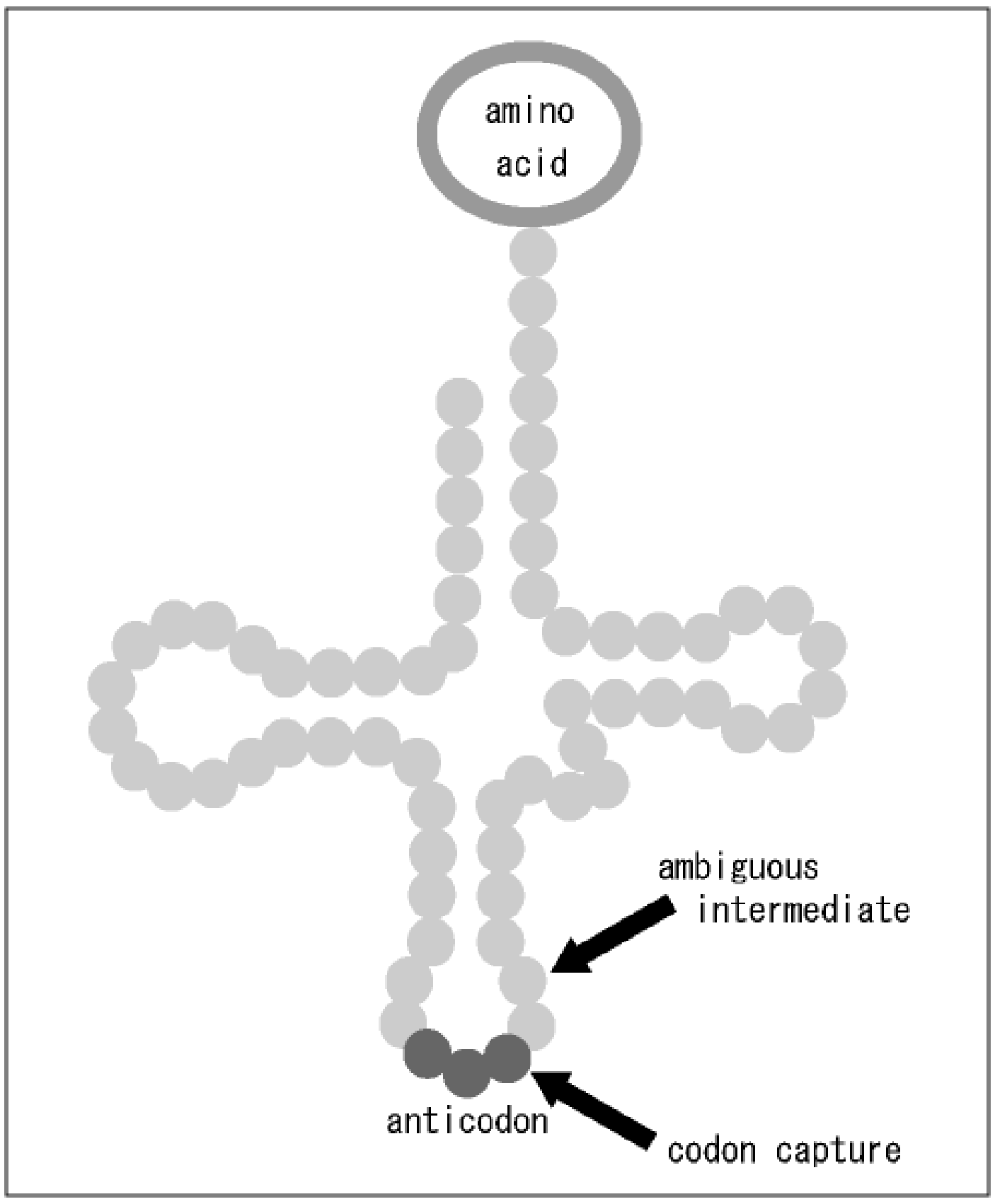

**Fig. 2.** Alterations in tRNA leading to the change in the range of codon recognition. Typically an alteration at the anticodon in tRNA is the trigger of the codon capture scenario, while the ambiguous intermediate scenario is triggered by that at locations other than the anticodon.

Another important basis of our model is the following discrimination[7] between two scenarios. The codon capture scenario is supported by observed codon reassignments from stop to an amino acid. On the other hand, the ambiguous intermediate scenario is supported by observed codon reassignments from one amino acid to another.

On these two bases we introduce a concept of codon level and discuss its structure as a key ingredient to obtain a unified framework for the codon capture and ambiguous intermediate scenarios.

## 2.2. Codon level

Here we model the affinity of tRNA for codon. The property of the target codon can be represented as a vector in a high-dimensional coordinate space. The coordinates constitute of the information of the properties of the codon such as the shape and the physical quantities. For example, the charge or hydrophobicity distribution of the codon is such a physical quantity. We name such a coordinate space, shape space, in short.

Since the purpose of this paper is to sketch our scheme, we adopt a scalar, representing the property of the target codon, instead of a vector, for simplicity. In the following we call this scalar the codon level. This simplified quantity is a key ingredient to obtain an insight about the mechanism of the genetic code evolution.

A similar construction[14] has been done in the study of immunity and a shape space is employed to describe the molecular recognition between antigens and antibodies. As the first step the 1-dimensional shape space has been thoroughly investigated there. In this paper we also focus on the 1-dimensional case as the first step of the study. Such a simplification reduces the reality of the model but we expect that the qualitative nature of the model is unchanged even for higher dimensional cases.

On the above-mentioned two bases we can naturally assume a structure in the distribution of the codon levels. The structure consists of intermittent scatter of clusters as shown in Fig. 3. The cluster corresponds to the family box of four codons in the genetic code table. These codons in the same family are often translated to the same amino acid by wobble base-paring between codons and anticodons. Such a wobble pairing is considered to result from two factors. One is the closeness of the codon levels in the cluster. The other is the spread of the range of the codon recognition of tRNA. This range is expressed as a section in the coordinate axis along the codon level as shown in Fig. 3. If a substructure with level gap exists in the cluster, codons separated by the gap may be translated to different amino acids. Here we have naively assumed a distinguishable distribution of the clusters and neglect an accidental overlap of clusters for simplicity. For the local part of codon levels where the genetic code evolution occurred in living cells, these assumptions were met so that the code evolved without confusion as seen later. Conversely the evolution is interrupted if these assumptions are not met.

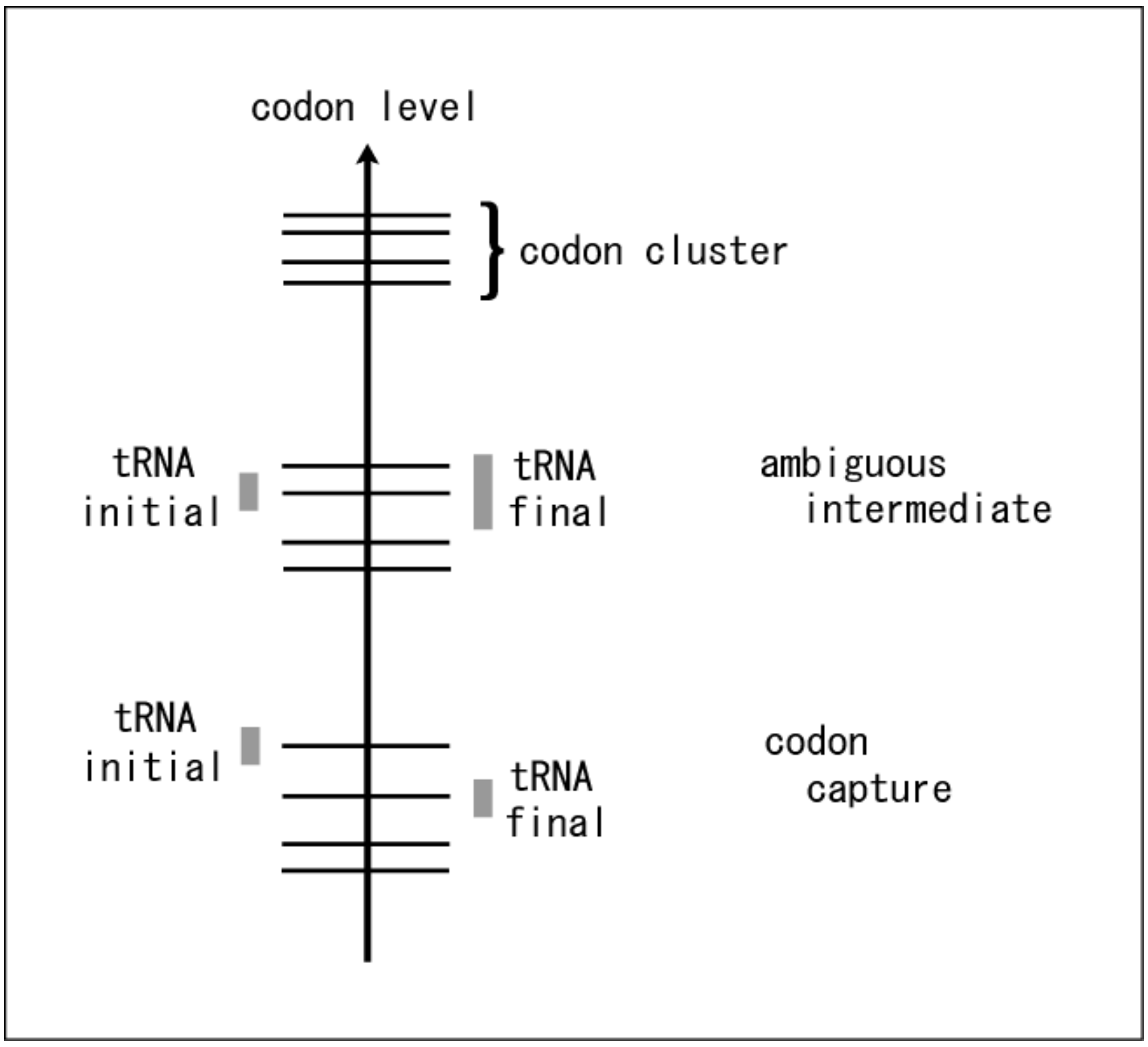

**Fig. 3.** Codon level scheme. In general the levels of four codons, belonging to the same family box of the genetic code table, form a cluster and clusters belonging to different families are separated by gaps in the level scheme. The range of the codon recognition of a tRNA is expressed as a section along the codon level axis. In the codon capture scenario the position of the center of the range jumps from a level to another level. In the ambiguous intermediate scenario the range spreads over two levels.

In our codon level scheme both codon capture and ambiguous intermediate scenarios are naturally understood as different evolutional route in the same shape space. While in the latter scenario the genetic code evolves only via intra-cluster transitions, in the former inter-cluster transitions also take part in. An alteration at anticodon in tRNA can be expressed as a jump between two codon levels which changes the codon-anticodon pairing. An alteration at locations other than the anticodon in tRNA can be expressed as a change in the width of the range of the codon recognition, since it does not change the codon-anticodon pairing directly. In the codon capture scenario the transition of the codons across the level gap between codon clusters also occurs as discussed later. This level scheme is consistent with the observation[3] that the ambiguous intermediate process only connects codons within the range of wobbling in a cluster. On the other hand, the process across the gap between the clusters is possible only in the codon

capture scenario.[3)]

2.3. Driving force of code evolution

So far we have discussed the trigger of the code evolution in a local manner where the affinity of tRNA for codon is the main subject. On the other hand, the selection of a new code is driven by global effects as genome size to be minimized, mutational pressure to minimize GC content in genomes and so on.[1)] In the following we take into account only representative effect of the code selection in evolution. The consideration on the other effects is left as an expanded study. In the case of the ambiguous intermediate scenario the selection is assumed to be driven by the cost of amino-acid synthesis according to the observation reported in ref. 7. Thus the code with more expensive amino acid is taken over by the code with less expensive one. In the case of the codon capture scenario[1)] cells which experience only neutral mutations survive against the natural selection.

## 3. Simulation

3.1. Simulation of ambiguous intermediate scenario

Here we perform an evolutional simulation on the basis of our codon level model focusing on the smallest part of the level scheme in which we can realize the ambiguous intermediate scenario. We pursue the time dependence of the tRNA distribution around two fixed codon levels in a codon cluster as shown in Fig. 4. In the initial stage codon-1 is recognized by tRNA-1 and codon-2 by tRNA-2. In the intermediate stage the ranges of the recognition of tRNAs spread by the mutational change and tRNA-1 recognizes both codon-1 and codon-2. In the final stage tRNA-2 is taken over by tRNA-1 by the cost-selection mechanism.

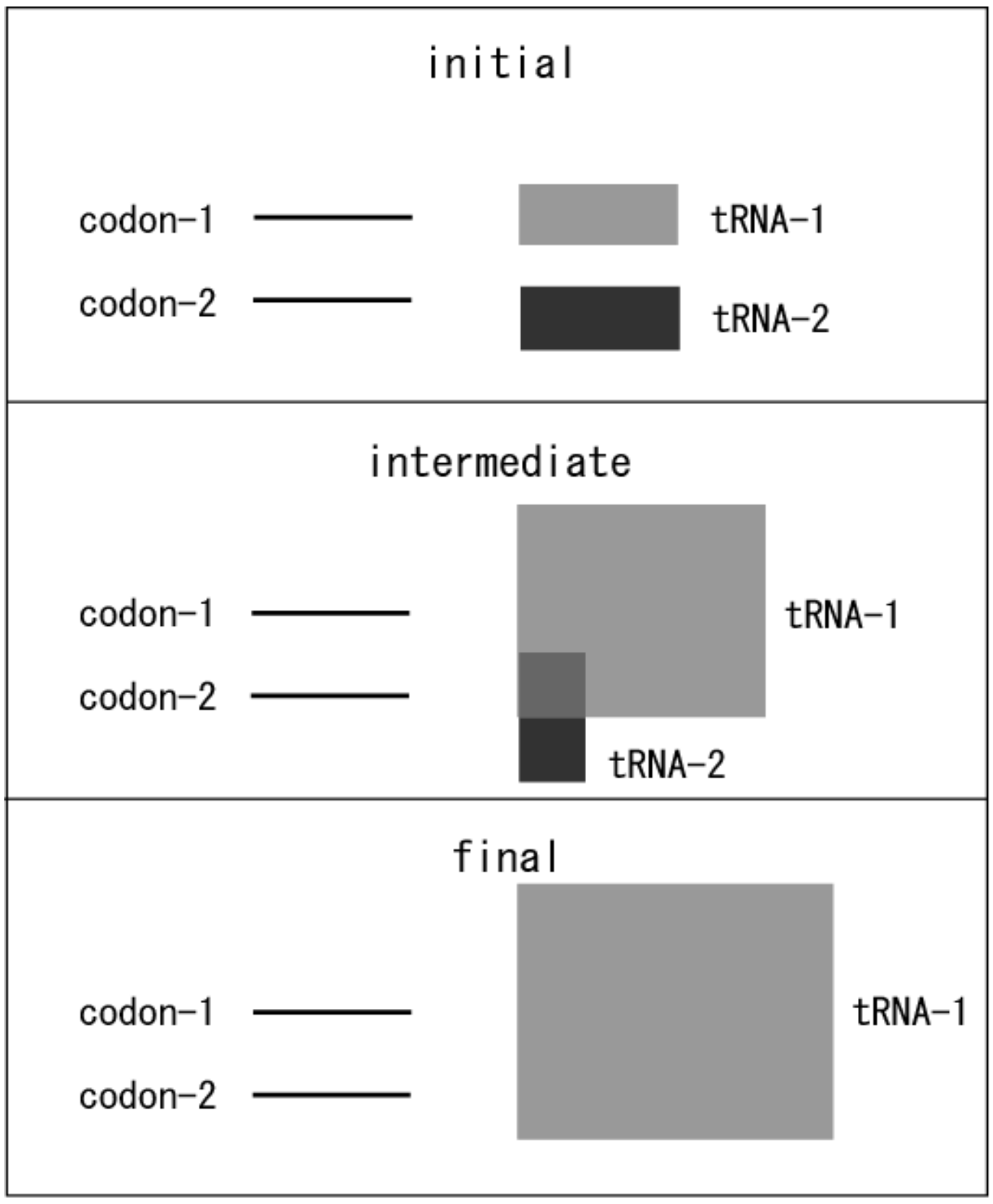

**Fig. 4.** Three stages in the ambiguous intermediate scenario. The height in the vertical direction represents the codon level. The range of the codon recognition of a tRNA is expressed as a section along the codon level axis. The population of a tRNA is expressed as a section in the horizontal direction. The ranges spread by mutations in tRNA. The populations changes according to the cost of the amino acid production and tRNA-2 is taken over by tRNA-1.

Our algorithm for the time evolution of tRNA in a cell is as follows. Step 0: We prepare the system consisting of two homogeneous groups of tRNAs. Step 1: We introduce alterations into tRNAs. Step 2: We eliminate ill-conditioned tRNAs with too narrow or too wide range of codon recognition. Step 3: We replicate tRNAs according to the cost of the synthesis of corresponding amino acids. Step 4: We divide the system into two, mimicking cell division, when the number of tRNAs reach a critical value. We repeat the procedure from Step 1 to Step 4 sequentially.

In the following we describe the detail of the simulation.

Step 0: As the initial condition for the simulation, the 0-th generation of the system, we prepare two homogeneous groups of tRNA, tRNA-1 and tRNA-2. The initial number of tRNAs of each group is $N_0$. The range of codon recognition for each tRNA, $t_i$, is given as $C_1 - W_1 \le t_i \le C_1 + W_1$ for tRNA-1 and $C_2 - \Delta - W_2 \le t_i \le C_2 - \Delta + W_2$ for tRNA-2 where each tRNA is labeled by number $i$. The codon level for codon-1 is $C_1$ and the center of $t_i$ for tRNA-1 is chosen as $C_1$. The codon level for codon-2 is $C_2$ and the center of $t_i$ for tRNA-2 is chosen as $C_2 - \Delta$. The values of the centers, $C_1$ and $C_2 - \Delta$, are assumed to be constants determined by the anticodons in tRNAs. The positive constant shift $\Delta$ represents the possibility that tRNA-2 may be taken over by tRNA-1. The widths of the range, $W_1$ and $W_2$, are constants at initial stage so that the groups of tRNA-1 and tRNA-2 are homogeneous.

Step 1: We introduce random alteration into randomly selected tRNA and give new width of the range, $w_i$, of codon recognition. Such an alteration is assumed to occur at locations other than the anticodons in tRNAs and change only widths of the recognition range keeping the center values unchanged. If an alteration occurs at the anticodon in tRNA, the center value is drastically changed and the recognition range is pushed out of the codon cluster now investigated. Such an alteration corresponding to the codon capture scenario is not considered here. The rate of the alteration, $\mu$, is fixed as $\mu = N / N_0$ where $N$ is the number of selected tRNAs for alteration. In order to represent the variety of nucleotides in tRNA we use a weight function $f(x) = w_0 \cdot [\frac{2}{L}(x - \frac{L}{2})]^{2\alpha+1}$ shown in Fig. 5 and alter $w_i$ into $w_i + f(x_i)$ where $x_i$ is a random number between 0 and $L$ generated for each tRNA. With this altered width $w_i$ new range of recognition, $t_i$, is given as $C_1 - w_i \le t_i \le C_1 + w_i$ for tRNA-1 and $C_2 - \Delta - w_i \le t_i \le C_2 - \Delta + w_i$ for tRNA-2. The weight function is introduced to represent numerically the change of the range of the recognition. The source of the change is not specified here among mutation of bases, base modification, RNA editing and so on. The index $\alpha$ regulates the time when the takeover between tRNAs begins. However, in order to save the simulation time we set the rate of the alteration $\mu$ as high as $10^{-2}$. The value of $\mu$ only sets the scale of the simulation time but has no realistic meaning. On the other hand, the transition time of the takeover is not affected by the choice of $f(x)$ as discussed later. The amplitude $w_0$ is chosen to allow alterations exceeding the range of ambiguous intermediate translation. Such highly altered tRNAs are eliminated in the next step.

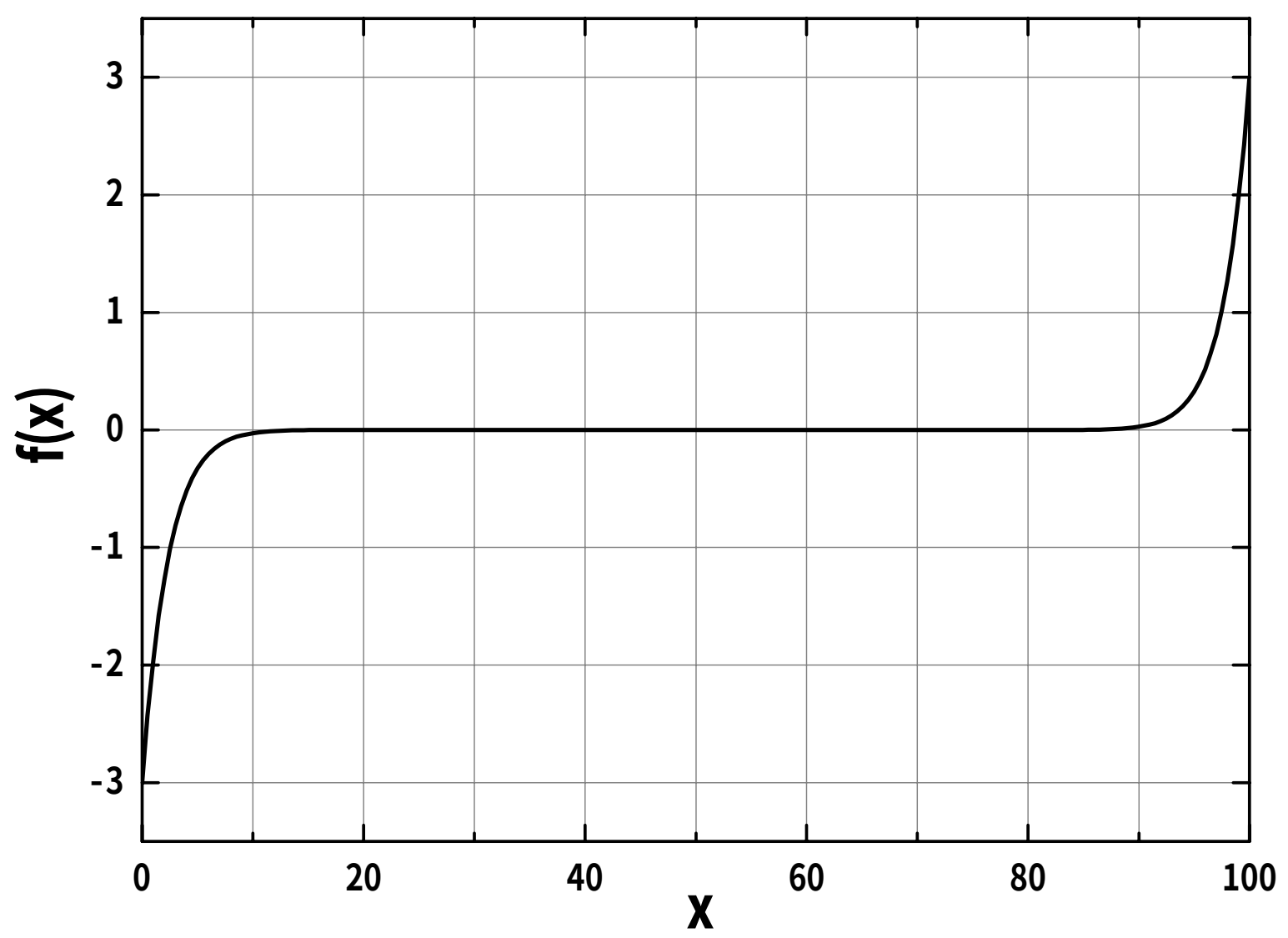

**Fig. 5.** Weight function $f(x) = w_0 \cdot [\frac{2}{L}(x - \frac{L}{2})]^{2\alpha+1}$ with $w_0 = 3$, $L = 100$ and $\alpha = 10$.

Step 2: We evaluate the effect of the alteration by keeping only tRNAs satisfying the condition $W_{\min} \le w_i \le W_{\max}$ which guarantees the ability of codon recognition. The lower value $W_{\min}$ represents a tolerance in order to function in some fluctuating environment not explicitly taken into account in our present model. The upper value $W_{\max}$ guarantees a singular nature of the recognition. The other tRNAs with poor recognition ability are eliminated.

Step 3: We replicate surviving tRNAs and expand the system. The rates of the replication, $r_1$ and $r_2$ , are assumed to be determined by the cost of the corresponding amino acid synthesis where amino acids transported by tRNA-1 and tRNA-2 are named aa-1 and aa-2, respectively. Thus we explicitly implement the selection among competing tRNAs. Since we focus on the case where tRNA-2 is taken over by tRNA-1, the rate, $r_1$, of the replication for aa-1 is set to be larger than that, $r_2$, for aa-2. The original tRNA-1s and tRNA-2s for the replication are chosen randomly under the constraint that the ratio of the total number of replication for tRNA-1 and that for tRNA-2 is $r_1 : r_2$.

Step 4: When the total number of tRNAs reach the critical value, $4N_0$, we make new systems at the next generation by dividing the system into two, each of which consists of randomly selected $2N_0$ tRNAs, and pursue one of the new systems afterward. The system of the $g$-th generation results from $g$ times divisions.

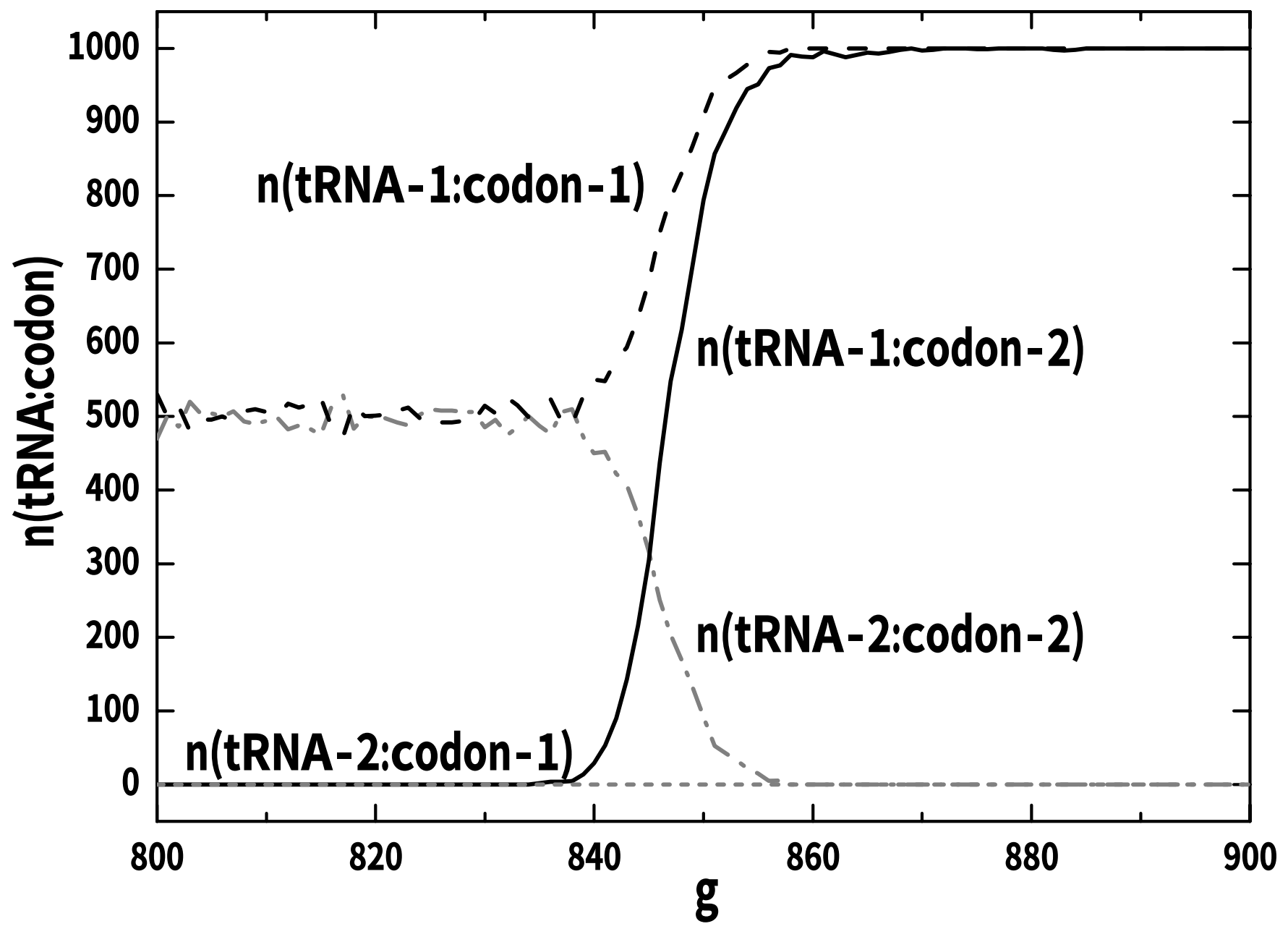

**Fig. 6.** Codon assignment transition for $N_0 = 500$, $C_1 = 15$, $C_2 = 10$, $\Delta = 1$, $W_{max} = 8$, $W_{min} = 0.1$, $\mu = 10^{-2}$. The number of tRNA is represented, for example, as *n*(tRNA-1:codon-2) which is the number of tRNA-1 recognizing codon-2. In the initial state all codon-1 are recognized by tRNA-1 and all codon-2 by tRNA-2. In the final state all of codon-1 and codon-2 are recognized by tRNA-1, since we have chosen as $r_1 / r_2 = 1.5$. Here $g$ represents the generation defined by the number of cell division.

In Fig. 6 an example of the time dependence of the codon recognition of tRNAs in a cell is shown. The tRNA with higher cost of amino acid synthesis, tRNA-2, is taken over by the other, tRNA-1. After the takeover both codon-1 and codon-2 are recognized by tRNA-1. Once a tRNA which recognizes both codons appears the takeover rapidly progresses irrespective of the rate. The rate controls the starting time of the takeover.

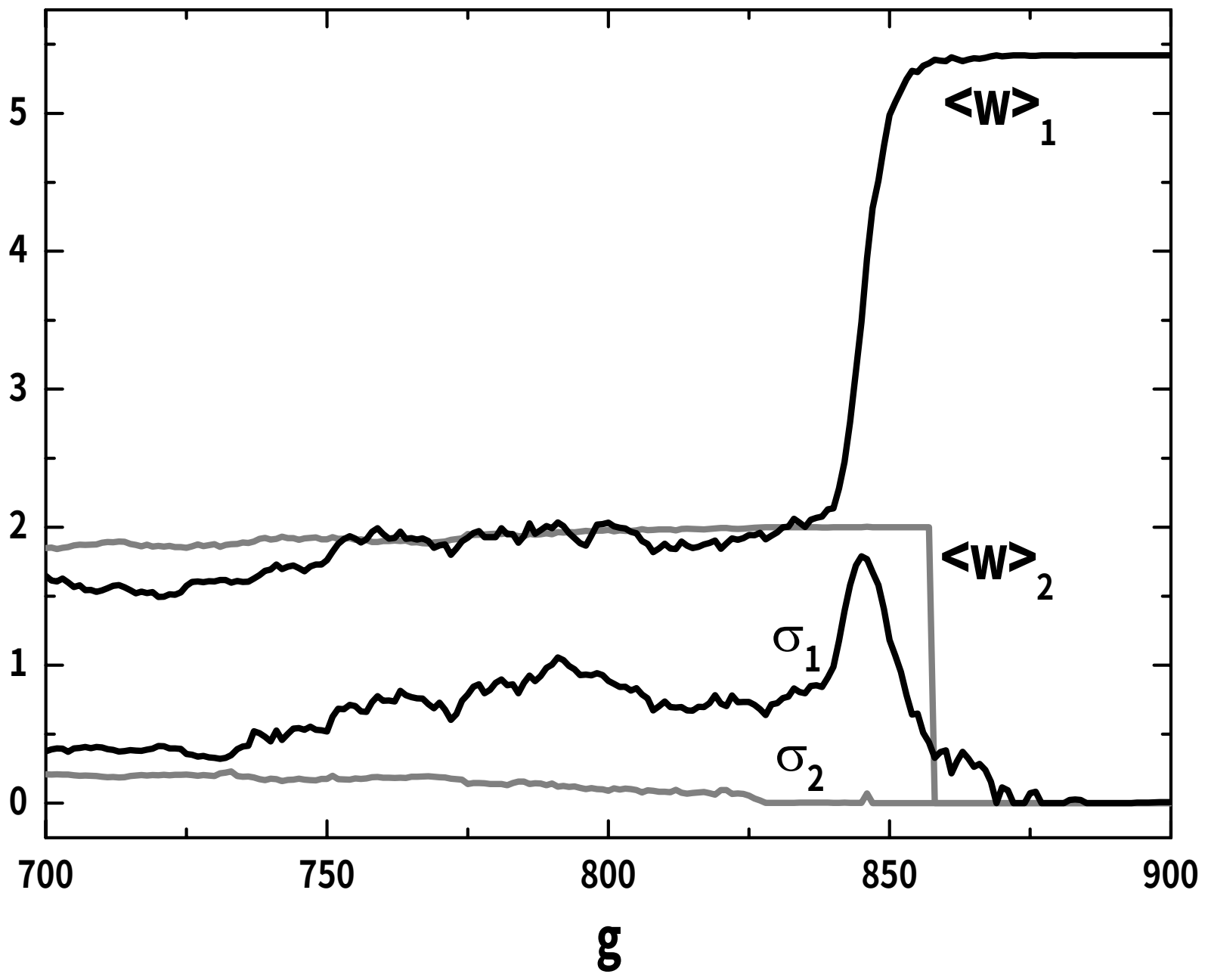

**Fig. 7.** Transition of the codon-recognition range for the simulation in Fig. 6. $\langle w \rangle_1$ is the average of $w_i$ over tRNA-1s. $\langle w \rangle_2$ is the average over tRNA-2s. $\sigma_1$ and $\sigma_2$ are the standard deviations around $\langle w \rangle_1$ and $\langle w \rangle_2$.

In Fig. 7 we show the transition in the width of the recognition ranges of tRNA-1 and tRNA-2. The fluctuation in the width of tRNA-1 increases toward the transition. On the other hand, the fluctuation of tRNA-2 is very small. After the transition tRNA-2s die out at $g = 858$ and the surviving tRNA-1s have almost the same width of $5.42$ with little fluctuation. While the mutational fluctuation is necessary for the transition, it calms down once the transition is accomplished.

The period of the ambiguous translation is independent of the cost of amino acid synthesis if the costs for two amino acids differ considerably as shown in Fig. 8. Since slight difference of the synthesis reactions leads to drastic change in the characteristic time of amino acid synthesis, it is taken for granted that the costs differ significantly when the takeover occurs. Thus the period, determined by an exponential growth of the difference in the numbers of tRNA-1 and tRNA-2, is about 10 generations for $N_0 = 500$ and very short in comparison with the time scale of evolution. Such rapidness of the takeover is the consequence of the exponential differentiation in the populations of tRNAs driven by the difference in the cost and reduces uncomfortableness of the non-unique translation of the code, by which two different kinds of amino acids result from a single codon, at intermediate state that is the weakness of the ambiguous intermediate scenario.

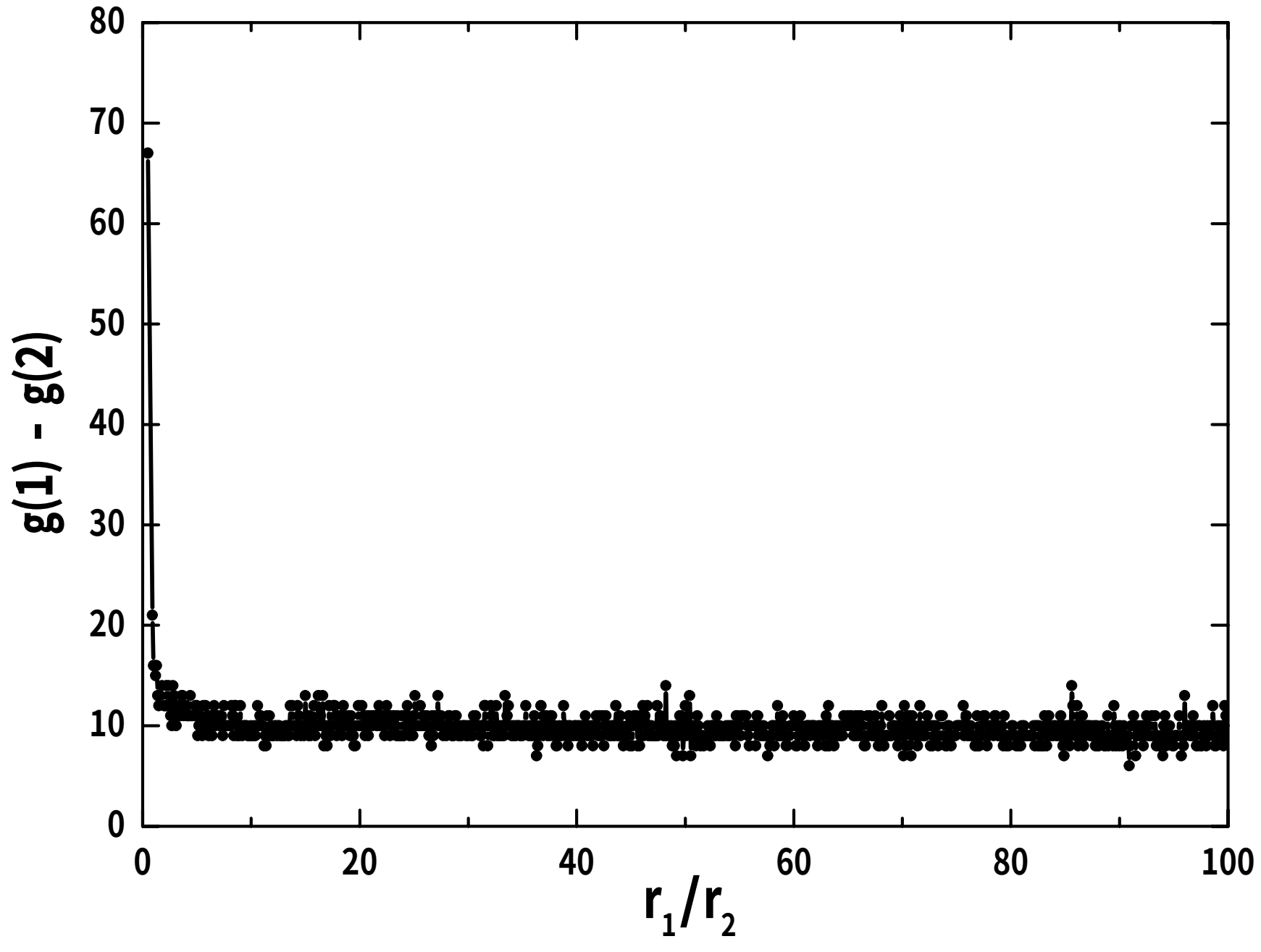

**Fig. 8.** Transition time as a function of $r_1 / r_2$ for the simulation in Fig. 6. At the generation $g(2)$, *n*(tRNA-1:codon-2) first exceeds 50 and at $g(1)$, it first exceeds 950. The transition time is estimated by $g(1) - g(2)$. The fluctuation of the data is due to the use of random numbers in the simulation.

The results of the simulation are insensitive to the choices of the other parameters. The value of $\Delta$ determines the direction of the takeover but does not control the transition time as shown in Table 1 in the case of the simulation in Fig. 6. While the functional form of the weight function $f(x)$ controls the starting time of the takeover, it does not affect the period so much as shown in Table 2 where the results are obtained using the same random number sequence as in the simulation of Fig. 6.

| $\Delta$ | $g(1)-g(2)$ | $g(2)$ |
|---|---|---|
| 0.5 | 14 | 841 |
| 1 | 14 | 841 |
| 1.5 | 14 | 842 |

**Table 1.** Neither the transition time $g(1)-g(2)$ nor the onset of the transition $g(2)$ is affected by the choice of the level shift $\Delta$.

| $\alpha$ | $w_0$ | $g(1)-g(2)$ | $g(2)$ |
|---|---|---|---|
| 6 | 3 | 16 | 9435 |
| 8 | 3 | 13 | 5738 |
| 10 | 3 | 14 | 841 |
| 10 | 4 | 15 | 409 |
| 10 | 5 | 12 | 33 |

**Table 2.** The transition time $g(1)-g(2)$ is hardly affected by changing the functional form of the weight function $f(x)$, i.e. changes in $\alpha$ and $w_0$, while the onset of the transition $g(2)$ shifts.

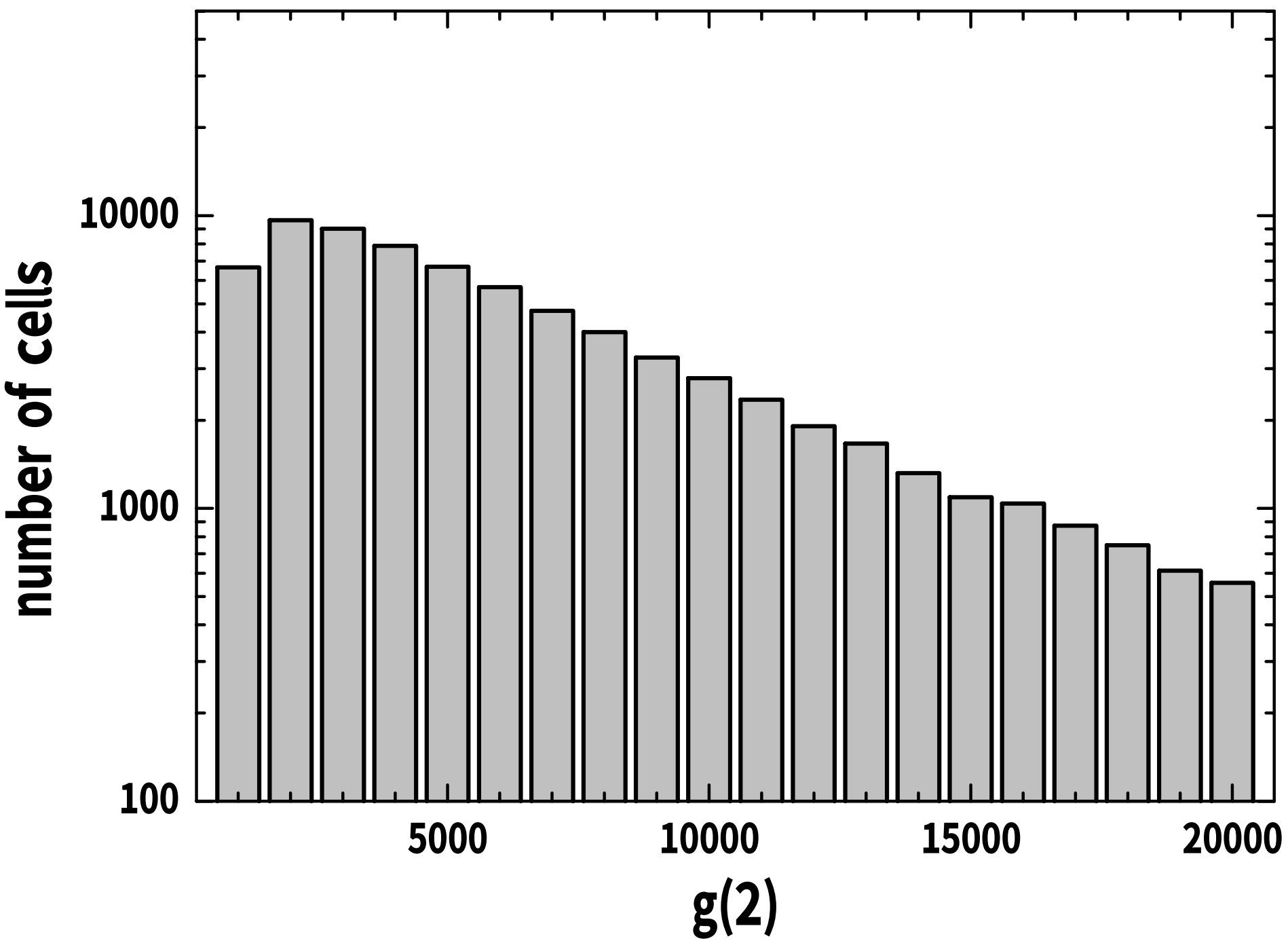

**Fig. 9.** Histogram of the onset of the transition $g(2)$. The simulation in Fig. 6 is repeated for $10^5$ independent random number sequences. Each sequence represents an independent cell.

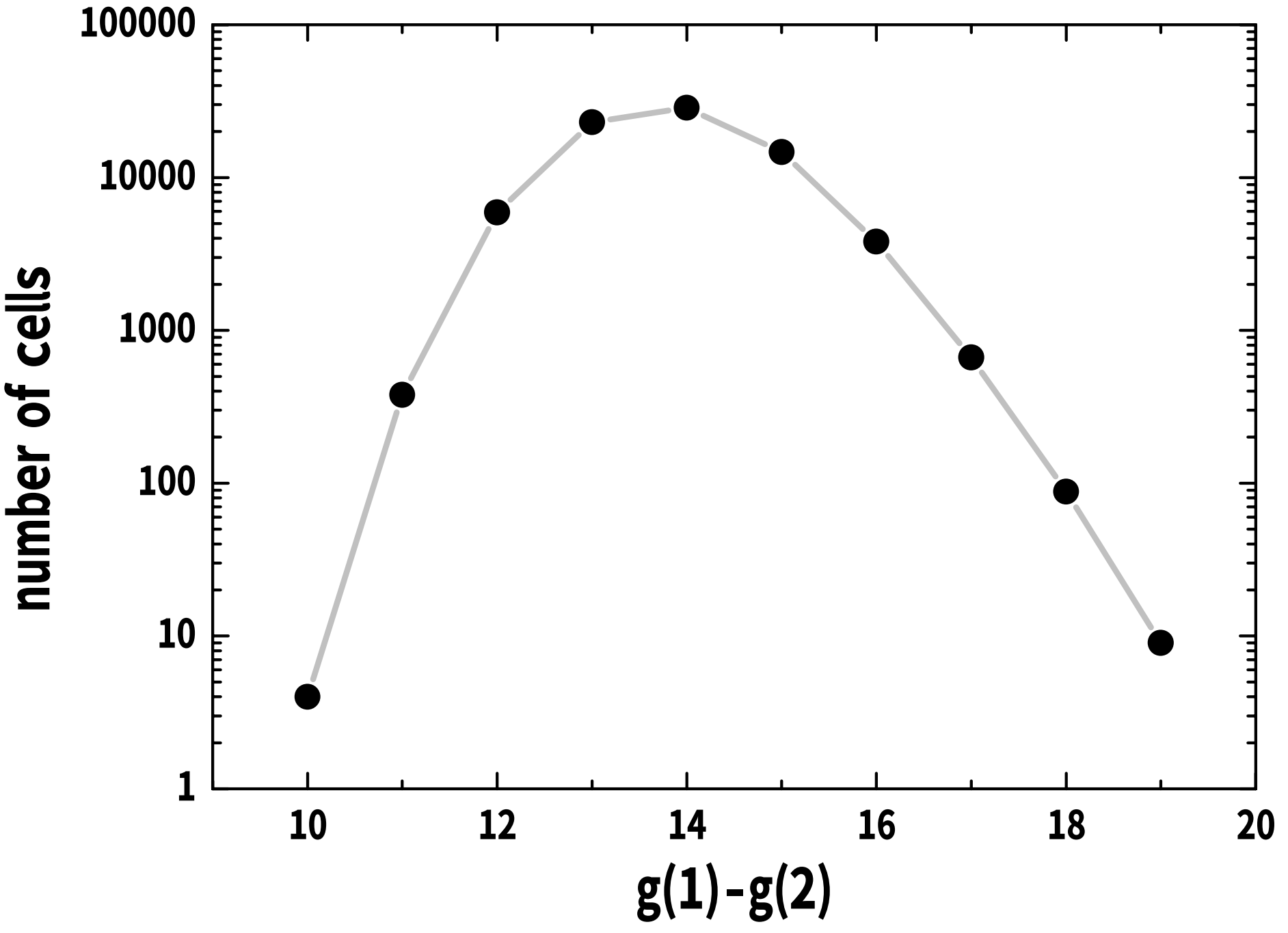

**Fig. 10.** Distribution of the transition time $g(1)-g(2)$ for cells appeared in Fig. 9.

Above results in this section are derived from a specific history of mutations, i.e. a sequence of random numbers. In Figs. 9 and 10 we show the results derived from $10^5$ independent sequences of random numbers. The transition begins by chance so that the onset $g(2)$ has no typical value and its distribution is exponential-like. Since the accumulation of mutations is needed for the onset, the number of cells for the smallest $g(2)$ is less than that expected from the exponential dependence. The transition time $g(1)-g(2)$ is controlled by the difference in the cost so that its typical value exists and the distribution is Gaussian-like. In short the transition begins probabilistically and progresses deterministically.

3.2. Simulation of codon capture scenario

Here we show a case of the simulation of the codon capture scenario. The global effects acting as the natural selection are taken into account as the rule of the simulation. As a typical case we focus on the capture of UGA-codon by a new tRNA introduced by the mutation of another tRNA. This case is described in detail in ref. 1. The processes considered in the simulation are shown in Fig. 11. In the following the driving force of the code evolution is assumed to be the mutational pressure replacing G by A and C by U. We perform an evolutional simulation starting with equivalent $N$ cells ($N = 10^5$) and evolving with discrete time $t$.

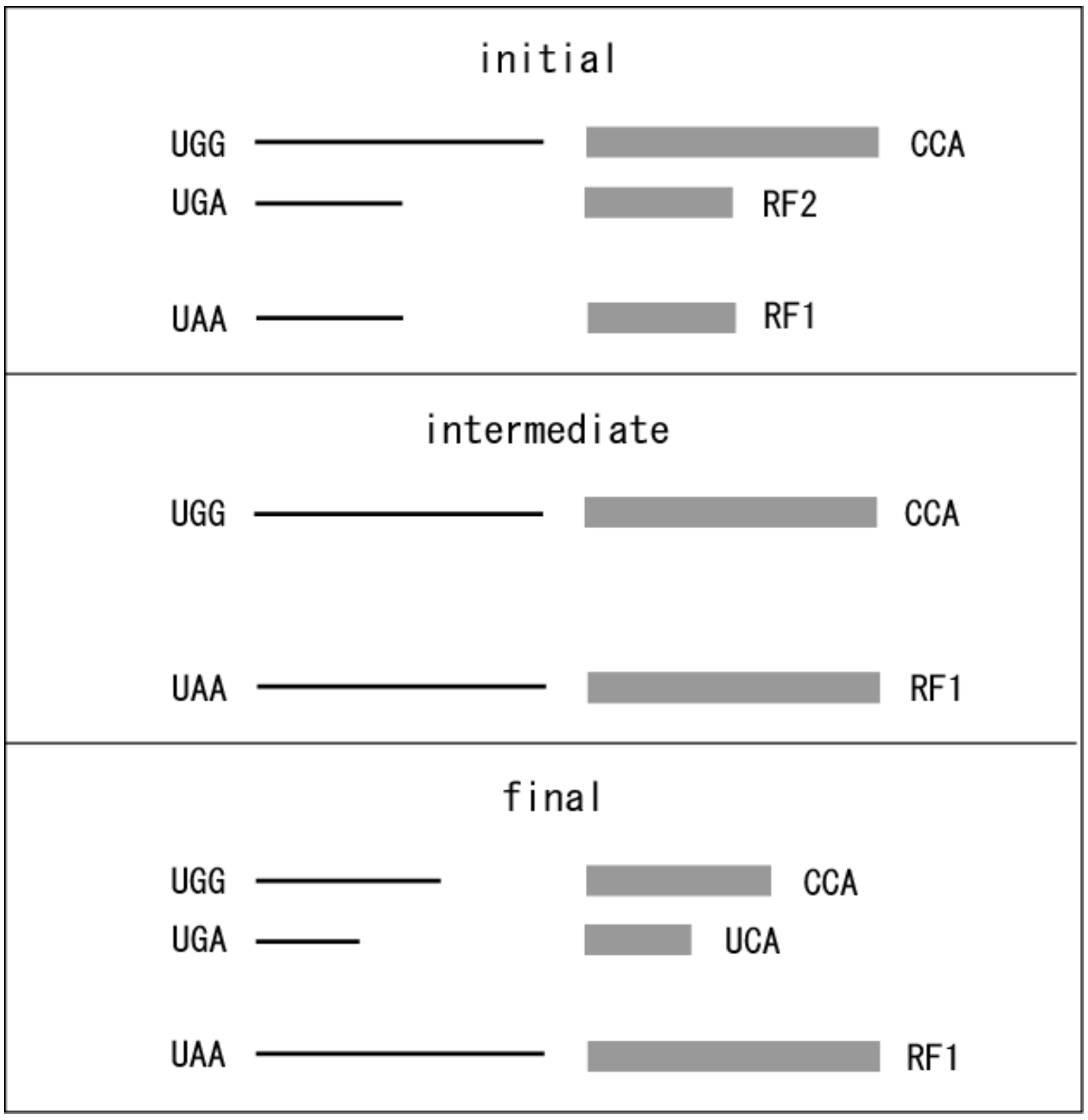

**Fig. 11.** Three stages in the codon capture scenario. The height in the vertical direction represents the codon level. The range of the codon recognition of a tRNA is expressed as a section along the codon level axis. The populations of codons and tRNAs are expressed as sections in the horizontal direction and change by point mutations in codons or anticodons.

In the initial stage UGA-codon is recognized by the release factor (RF2) and acts as a stop codon. Under the mutational pressure the codons are changed as UGG → UGA → UAA and the rates of the mutations are assumed to be equal to $\mu = 10^{-3}$ for each codon at every time step. Another point mutation, UGG → UAG, is also implemented in the simulation by the same rate. UGG-codon is recognized by the tRNA with CCA-anticodon and translated into the amino acid, tryptophan (Trp). UAA-codon is recognized by the release factor (RF1) and acts as a stop codon. UGA and UGG belong to the same codon cluster but UAA belongs to another cluster. CCA-anticodon is also assumed to change by the point mutation as CCA → UCA or CCA → CUA by the same rate.

A mutational change is accepted, if it does not change the output of the translation. Namely only neutral changes are accepted.[1] In this respect only UGA → UAA is accepted among above mutational changes. All the other mutational changes, UGG → UGA, UGG → UAG, CCA → UCA and CCA → CUA, are rejected, since the change in the translation is assumed to cause a fatal error to the cell activity. Then cells suffered by the fatal error are dead and the other cells survive when the mutations are accumulated.

For simplicity we do not implement the behavior of the release factors, RF1 and RF2, but the populations of these are assumed to evolutionally change according to the populations of UGA- and UAA-codons in the same cell.

At the intermediate stage which is the turning point between the initial and the final stages, UGA-codon and its corresponding release factor RF2 vanish. After this point all the stops for the translation are governed by UAA-codon and its corresponding release factor RF1.

In the final stage the mutational change, CCA → UCA, is possible, since the role of UGA-codon, which is recognized by tRNA with UCA-anticodon, is open. If tRNA with CCA-anticodon and tRNA with UCA-anticodon bring the same amino acid, the mutational change, UGG → UGA, is accepted, since it does not change the translation. Thus UGA-codon is captured by tRNA with UCA-anticodon and the role of UGA-codon in the translation changes from stop to the amino acid Trp. Other changes by point mutations, UGG → UAG, UGA → UAA, CCA → CUA and UCA → UUA, are fatal and rejected.

The criterion for the acceptance or the rejection is employed for simplicity and the tolerance, for example resulting from the resemblance among amino acids in their physical quantity, is not taken into account also in ref. 1.

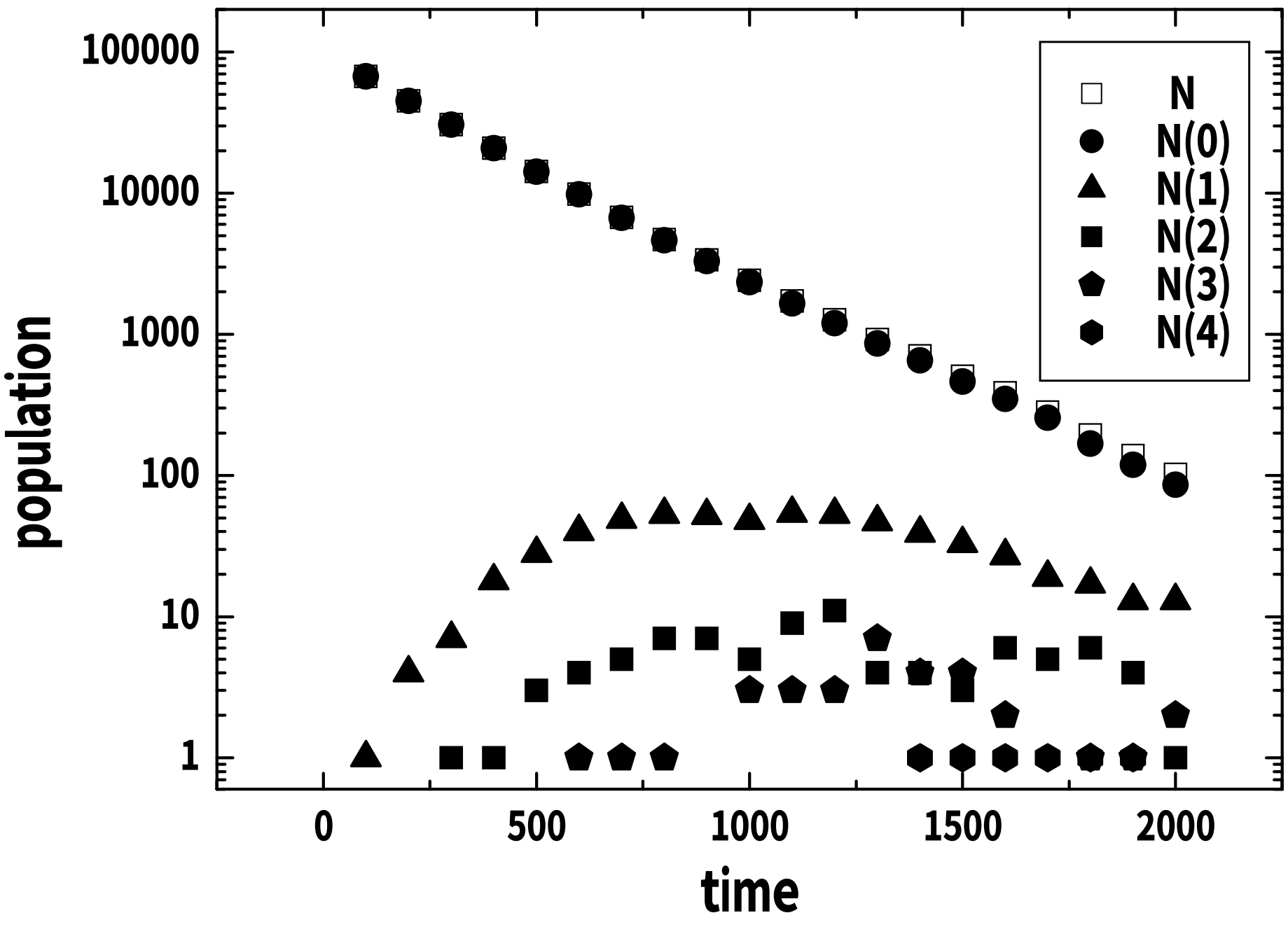

**Fig. 12.** The populations of cells as a function of time step $t$ for $N = 10^5$ at $t = 0$ and $\mu = 10^{-3}$. $N(i)$ represents the population of cells where $i$ is the number of UGA-codon captured by tRNA with UCA-anticodon in a cell. $N$ represents the total population of surviving cells and $N = \sum_{i=0}^{4} N(i)$.

At the start of the simulation each cell has 4 UGG-codons, 2 UGA-codons, 2 UAA-codons and 4 tRNAs with CCA-anticodon. The result of this small simulation is shown in Fig. 12. $N(i)$ is the population of cells where $i$ is the number of UGA-codon captured by tRNA with UCA-anticodon in a cell. Thus non-zero $N(i)$ for non-zero $i$ represents the population of the cells in which the code evolved by the codon capture process.

Since we do not implement the growth and the division of cells, the population of surviving cells $N$ decreases as the simulation time step $t$ increases. If we implement the exponential growth of the cell population, we can continue the simulation with considerable population of surviving cells.

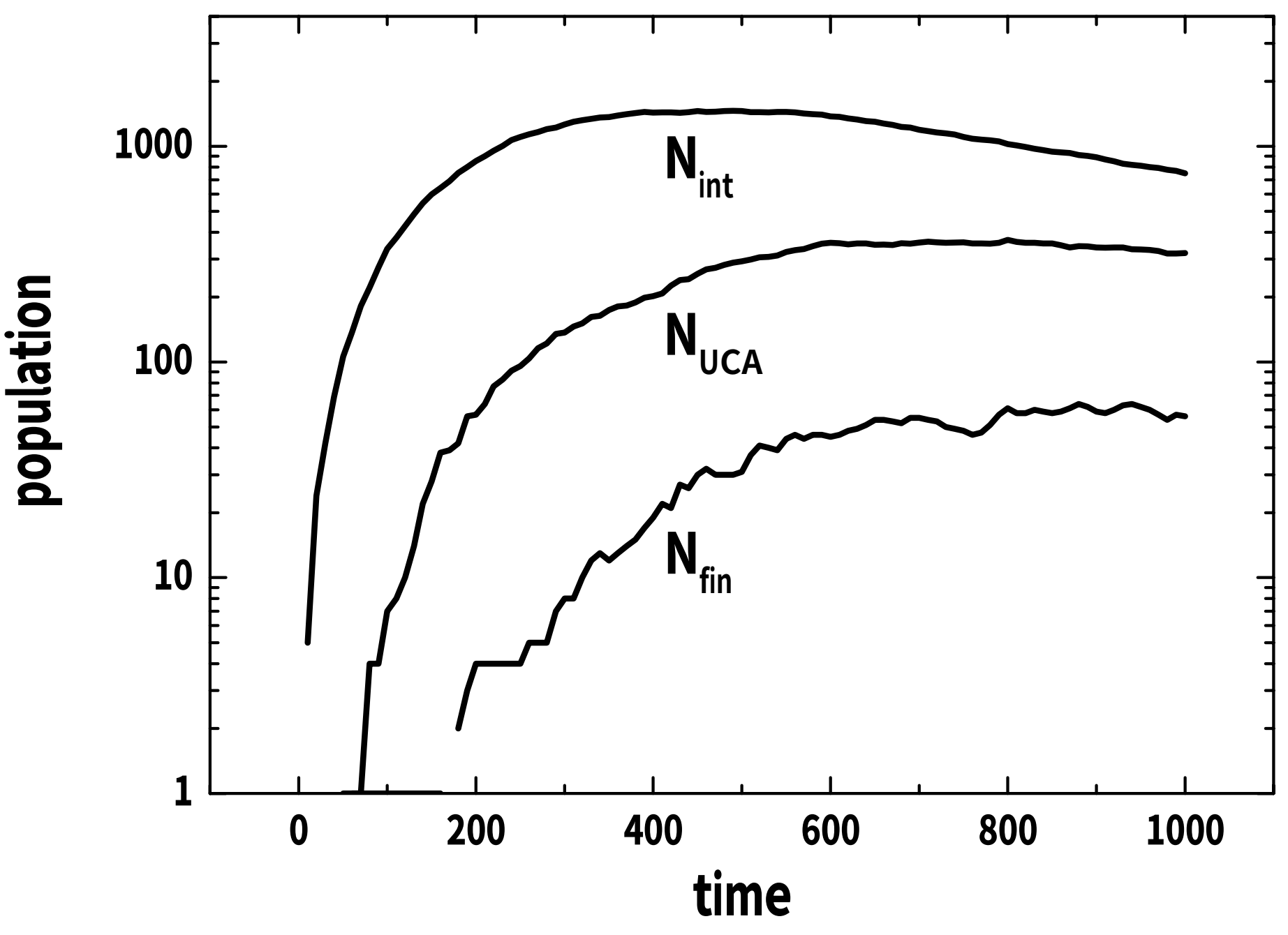

**Fig. 13.** The populations of cells as a function of time step for the simulation in Fig. 12.

In Fig. 13 we show the sequence of the codon capture transition. $N_{\rm int}$ and $N_{fin}$ represent the population of cells in the intermediate and final stages, respectively. $N_{UCA}$ represents the population of cells with UCA-anticodons. The development of $N_{fin}$ follows that of $N_{UCA}$, because the set of the former cells is the subset of the latter. The relation between $N_{UCA}$ and $N_{\rm int}$ is the same.

## 4. Conclusion

We have given a physical description of the genetic code evolution making use of the shape space of codons and tRNAs. In the framework of our simplified model the codon capture and ambiguous intermediate scenarios are described in a unified manner. In terms of the codon levels two scenarios are classified into two different routes of the evolutional process.

Our simulation implemented cost selection of amino acids has demonstrated a rapid transition of the code change in the case of the ambiguous intermediate scenario. This rapidness of the transition covers the drawback, the non-unique translation of the code during the transition, of the scenario. It should be noted, however, that the cost is only one of the selection pressures and for more realistic description we have to consider the selection from more global point of view for the environment of cells.

In the case of the codon capture scenario it is demonstrated that cells with only neutral mutations survive and the code can change only in such cells.

Although our unified description in the shape space has made some progress in understanding the genetic code evolution, it has only integrated the scenarios already reported and the simulations have given us only expected results. Starting from the present one we have to develop the model with the ability to predict some non-trivial results as the next stage of investigations.

Our demonstrations in both cases are too simple and abstract and we should employ more realistic model in order to describe actual phenomena. Such an elaboration is left to an extended version of the present model.